# Simulating Bloch points using micromagnetic and Heisenberg models

Thomas Brian Winkler[1*], Marijan Beg[2], Martin Lang[3,4], Mathias Kläui[1*], Hans Fangohr[3,4*]

*1 - Institute of Physics, Johannes Gutenberg University, Staudinger Weg 7, 55128 Mainz, Germany*

*2 - Department of Earth Science and Engineering, Imperial College London, London SW7 2AZ, United Kingdom*

*3 - Faculty of Engineering and Physical Sciences, University of Southampton, University Road, Southampton SO17 1BK, United Kingdom*

*4 - Max Planck Institute for the Structure and Dynamics of Matter Hamburg, Luruper Chaussee 149, 22761 Hamburg, Germany*

* Email: twinkler@uni-mainz.de, klaeui@uni-mainz.de, hans.fangohr@mpsd.mpg.de

## Abstract

Magnetic Bloch points (BPs) are highly confined magnetization configurations, that often occur in transient spin dynamics processes. However, opposing chiralities of adjacent layers, for instance in a FeGe bilayer stack, can stabilize such magnetic BPs at the layer interface. These BPs configurations are metastable and consist of two coupled vortices (one in each layer) with the same circulation and opposite polarization. Each vortex is stabilized by opposite sign Dzyaloshinskii-Moriya interactions. An open question, from a methodological point of view, is whether the Heisenberg (HB) model approach (atomistic model) is to be used to study such systems or if the – computationally more efficient – micromagnetic (MM) models can be used and still obtain robust results. We are modelling and comparing the energetics and dynamics of a stable BP obtained using both HB and MM approaches. We find that an MM description of a stable BP leads qualitatively to the same results as the HB description, and that an appropriate mesh discretization plays a more important role than the chosen model. Further, we study the dynamics by shifting the BP with an applied in-plane field and investigating the relaxation after switching the field off abruptly. The precessional motion of coupled vortices in a BP state can be drastically reduced compared to a classical vortex, which may also be an interesting feature for fast and efficient devices. A recent study has shown that a bilayer stack hosting BPs can be used to retain information [1].

## Introduction

Studying magnetic thin-film and multilayer-stack systems on the mesoscopic scale is of fundamental interest to researchers in basic science as well as for applications. On the one hand, the investigation of topologically protected spin structures, such as Skyrmions [2] or Hopfions [3], and their manipulation with fields and spin-currents allows for analyzing the dynamics of topological quasi-particles as well as spin phenomena. On the other hand, the realization of low-power and high-density next-generation storage devices and implementation of neuromorphic computing schemes into such systems is of economic relevance and great ecological interest [4–6].

The Dzyaloshinskii-Moriya interaction (DMI) plays a key role in the stabilization of magnetic solitons [7]. This indirect, asymmetric exchange interaction favors spin canting between neighboring spins and gives rise to a certain chirality and topologically stabilized spin structures [2].

In our investigation, we focus on the most highly confined soliton, a Bloch point (BP) [1,8–10]. A BP is a magnetic singularity, a point where the magnetization $\boldsymbol{m}$ diverges. BPs commonly occur in transient spin dynamics processes [11–13]. They are usually not stable as static spin structures in thin films. However, they have been found to be (meta-)stable in simulations and experiments [14,15]. In a simulation-based study, Beg et al. [8] were able to show that a BP can be stabilized by stacking on top of each other two FeGe layers with opposite sign of the (bulk) DM energy constant $D$. A magnetic vortex forms in each layer. The two vortices – one in each layer – have identical circulation and opposite polarization to satisfy the respective material constant $D$ (and the associated chirality) of their hosting layer.  A vortex typically has four energetically equal states, defined by a combination of the circulation $c$ (±1) and polarization $p$(±1).  In a material with DMI, two combinations of $c$ and $p$ are energetically more favorable (the chirality of DMI couples circulation and polarization). Stacking two layers with DMI strengths $D_1$ and $D_2 = -D_1$, the DMI energy can stabilize two vortices with same circulation but different polarization, leading to a BP at the interface between the top ($D_1$) and bottom ($D_2$). So, the preferred handedness in each layer is established at the cost of one magnetic singularity. In our case, we find a Head-to-Head BP (HH-BP), an illustration is shown in *Fig. 1a*. We use the terminology HH-BP according to Beg et al. [8], while in the nomenclature of Malozemoff and Slonczewski [13] the structure would be named *converging circulating BP*.

It is unclear to what extent a BP can be investigated within a discretized micromagnetic model (MM) approach, due to the strong spatial variations of the magnetization field [11,16]. MM results may not be robust when the spin canting between neighboring simulation cells increases above a certain angle, for both finite-element and finite-difference approaches. Atomistic or multi-scale simulations [11,16–18] are better suited to investigate such systems. However, theoretical and numerical investigations have been previously carried out to analyze a BP within the MM framework [8,19–22]. In Ref. [19], the authors have shown that the existence of a Bloch point during a vortex core reversal in an MM framework is possible. The calculated energy barriers depend on the mesh discretization, indicating potential issues with this approach. In Ref. [23] a stable BP was found experimentally by X-ray vector nanotomography. While in principle DMI is not strictly necessary to stabilize such structures, it supports the nucleation of a BP in our case by lifting the degeneracy of handedness of the vortex configurations and fixes the BP at the interface. In Ref. [20], the authors were able to predict the existence of chiral bobbers in an MM framework, a stable spin structure containing a BP. Also, the existence of “dipole strings” was shown in recent years, a magnetization configuration containing two BPs [24,25]. A. Savchenko *et al.* showed that the existence of the Dipole string in an FeGe cylinder [24] can be obtained in simulations by modelling fabrication defects, e.g., switching off the DMI at the surface of the cylinder. This is related to our approach of inducing BPs by modelling different regions with different DMI strengths.

A more accurate model than the MM model to describe such a system with spatially strongly varying magnetization is the Heisenberg (HB) model, which allows one to simulate every magnetic unit cell with its interactions. The HB model inherently can deal with arbitrary variations of the spin structure. The MM model is derived from the HB model and assumes a spatially slowly varying magnetization. The drawback of the HB model is the high computational effort since every magnetic unit cell is simulated separately, while in an MM framework a system discretization can be chosen that requires up to several orders of magnitude fewer simulation cells than the HB model (depending on the magnetic parameters).

In this work, we systematically study the energetics and dynamics of a stable BP in a bilayer stack within the MM framework and compare the results with the more accurate HB model. A key parameter for us is the mesh discretization $d$, which we vary from a few $\mathrm{nm}$ down to a few $\mathrm{\AA}$. We on purpose also choose discretizations $d$ smaller than the atomic lattice constant $a = 0.4679\ \mathrm{nm}$ for FeGe, to probe the MM framework. For a detailed discussion of the fundamental differences between the MM and the HB model in our implementation, we refer to Ref. [11].

The study consists of two parts: First, we nucleate a stable Bloch point according to [8] and calculate the energy obtained for different discretizations. As a comparison, we also nucleate a vortex in the bilayer system. This "bilayer vortex" (BV) – a vortex extending over both layers with same polarization and circulation in both layers – enables us to compare the BP configuration's behavior with that of a configuration that is generally accepted to be accurately described within the MM model. *Fig. 1* illustrates the BP state and the BV state. The in-plane rotation (circulation) is the same in both structures. However, the BP state shows two vortex cores with different polarizations, generating the BP at the interface, while the BV state shows one vortex core pointing upwards. The vortex core size variation comes from the different DMI sign in the two layers, leading to favorable (thicker) or unfavorable (thinner) circulation compared to the respective DMI handedness. To minimize shape approximation errors in the finite-difference method, we use a square-shaped nano-scale geometry.

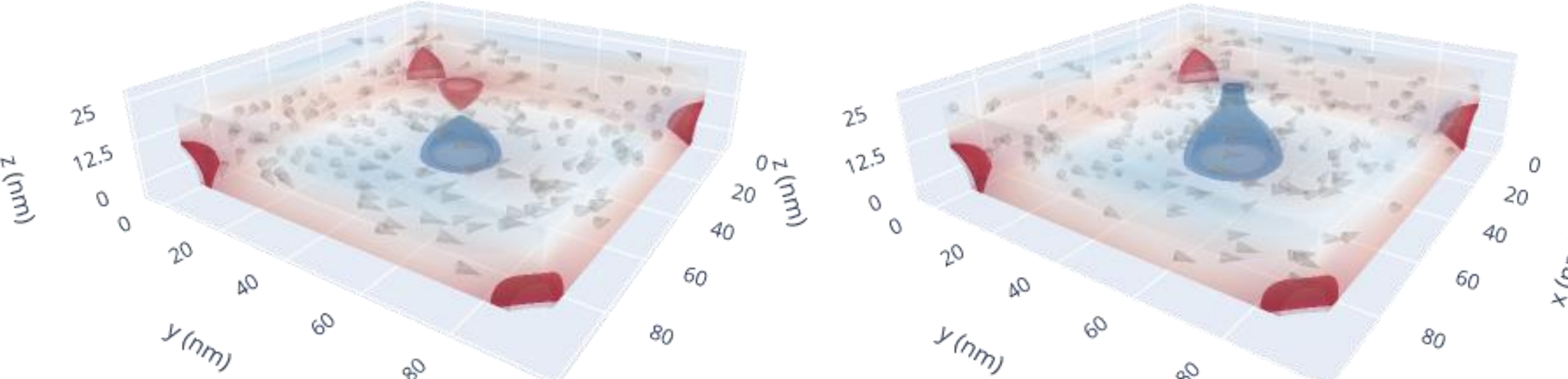

***Figure 1:*** *Two of the investigated structures in a square-shaped geometry. BP state (left) and bilayer vortex state (right) are both stable in the bilayer stack. The iso-surfaces show magnetization with* $m_z$ *< - 0.85 (red) or* $m_z$ *> 0.85 (blue). Grey cones visualize the in-plane magnetization. Not every simulation cell is plotted. Due to the rectangular shape and the DMI, the bottom corners also show a large OOP component.*

In the second part, we move the stable BP with an external field applied in-plane (IP) with a strength of $25\mathrm{mT}$. We then switch off the field and analyze the relaxation process of the BP. We compare the dynamics with a "classic" magnetic vortex without DMI (all other material parameters based on FeGe) and the vortex in the bilayer-stack. Again, we vary the mesh discretization to probe the MM framework. For this part of the study, a circular-shaped nano-disk is chosen to make results of the vortex motion more comparable with analytical solutions or calculations based on the Thiele model [26]. Bloch point motion was already investigated in earlier studies. Results were often compared quasi-statically [14,27], or treated analytically [13]. A direct comparison of the MM and the HB model for a BP-mediated Skyrmion tube annihilation reveals a discrepancy in the maximum velocity a BP can reach [28]. Furthermore, a strong cell-size dependence was found. We are particularly interested in the numerical accuracy of a stable BP in the MM framework, and when a transition to a more precise HB model is necessary. As MM simulations at finite size are core drivers of application development in the area of spintronics, BPs are becoming more often the focus of quasi-particle investigation, e.g., as part of chiral bobbers [29]. We recognize an interest in obtaining reliable and robust results with MM simulations with finite-sized meshes [30].

# Methods

Our simulation-based study is mainly done with *MicMag2* [31], an expanded version of MicroMagnum [32,33], a free, open-source simulation package, that allows for fast GPU-accelerated time integration with double-precision. The fast calculation of the magnetostatic energy is done with the FFTW package [34]. However, many in-house extensions have been made in recent years [11,18,35] to adapt the code to today´s research interests, such as the DMI interaction. In particular, the HB model was included to allow for accurate simulation of spin structures that are normally outside the scope of the MM framework. *MicMag2* uses the finite-difference method and the Runge-Kutta evolver of fifth order for the time integration [36].

The study is considering an FeGe bilayer-stack, and we adopt all material parameters from [8]. The evaluated energy contributions are the exchange energy ($A = 8.78\,\mathrm{pJ/m}$), the demagnetization field ($M_\mathrm{s} = 0.384\,\mathrm{MA/m}$) and the DMI ($D_1 = -D_2 = 1.58\,\mathrm{mJ/m^2}$) with opposite sign in each of the layers. The thicker bottom layer ($D_1 > 0$) has a thickness of $t_\mathrm{bottom} = 30 \cdot a$, the thinner top layer ($D_1 < 0$) a thickness of $t_\mathrm{bottom} = 20 \cdot a$, resulting in a 3/2 ratio, enabling us to vary the MM discretization in reasonably many steps without distracting the ratio of the layers in the MM framework. The side-length $l$ of the square-shaped geometry was chosen as $l = 200 \cdot a$. For the coupling between the two layers, we consider direct exchange and dipolar effects.

To move the BP, the bilayer vortex and the classical vortex in the respective simulated samples out of equilibrium, we use an external magnetic field. To nucleate the BP, we initialize the system as homogeneously magnetized in the $z$-direction. Then we let the system evolve according to the Landau-Lifshitz-Gilbert-equation [37,38]. To nucleate the vortex in the bilayer stack, we also initialize the system in the $z$ direction and then relax with the dissipative LLG equation (without the precessional term). For nucleating the classical magnetic vortex, we initialize the system with an analytical expression that approximates a vortex configuration and then let the system evolve. All relaxations are performed with a stopping criterion of a maximum torque of τ = $0.5\,\mathrm{degree/ns}$. The important length scales for choosing the discretization are the exchange length $l_{\mathrm{ex}} = \sqrt{(2\,A/\mu_0\,M_S^2)} = 9.73\ \mathrm{nm}$, the helical length $L_{\mathrm{D}} = \frac{4\pi A}{D} = 70\ \mathrm{nm}$ and the atomic lattice constant of FeGe, $a = 0.4679\ \mathrm{nm}$. We choose MM discretizations below $l_{\mathrm{ex}}$ and even below $a$ to probe the stability of the BP in the MM framework.

# Results and Discussion

## Energy of the BP state

Beg et al. have shown that a stable BP can be nucleated in an MM framework [8]. We repeat the simulation for different cell sizes in the MM framework, as well as in the HB model. We compare the energy of the stable BP with a bilayer vortex in the same system. The latter configuration would widely be regarded as a structure that can be modeled using micromagnetics, and we use it as a comparison system to provide context for the observed inaccuracies.

According to Bogdanov et al. [39], vortices in systems with induced DMI are only stable up to some critical DMI strength $D > 4/\pi \cdot D_{crit}$ with $D_{crit} = \sqrt{A \cdot K}$ and $K = \frac{\mu_0}{2} M_S^2$ the effective shape anisotropy. It is worth mentioning that we are using $D \approx 1.75 \cdot D_{crit}$ in our bilayer systems and still find vortex-like structures.

*Fig. 2a* shows the total energy, and the single energy terms for our studied system and the comparator systems on the y-axis. On the x-axes, we vary the MM discretization between $d_{\mathrm{max}} = 10 \cdot a = 4.679\ \mathrm{nm}$ and $d_{\mathrm{min}} \approx 0.2599\ \mathrm{nm}$. We note that the BP configuration has a lower energy than the BV configuration, supported by the strong DMI value chosen here. From *Fig. 2d*, we see that the DMI energy in the BP state is minimized compared to the vortex state. As the material has two DMI constants of opposite sign, the BV state has the less favored handedness in one of the layers, whereas the Bloch point configuration minimizes exactly that energy by aligning the handedness of the vortices in bottom and top layer with the preferred handedness of the DMI material constants. *Fig. 2b* shows that the exchange energy of the BP configuration is higher than that of the BV configuration. The Bloch point itself contributes to an increase in the exchange energy.

The presence of the Bloch point also increases the magnetostatic energy density a little (*Fig. 2c*). Overall, the energy reduction in the DMI energy density for the BP configuration outweighs the increments of energy for the magnetostatic and the exchange energy. This is in line with Ref. [8] which reports the BP as the stable configuration with the lowest energy.
We note that for the largest chosen MM cell size of $\mathrm{d_{MM}} = 10 \cdot \mathrm{a} = 4.679\ \mathrm{nm}$ the bilayer vortex state is not stable, but only the BP state exists in equilibrium.

The total energy is decreasing in both systems with decreasing cell size (Fig. 2a). This is reasonable since the circular magnetization in the vortices can align better in a finer grid, leading to less spin canting between adjacent cells. It is noteworthy that the total energy of the BV state varies more with the discretization than the BP state, although less spin canting between cells is observed. The magnetostatic energy is saturating faster for smaller fields (Fig. 2d).

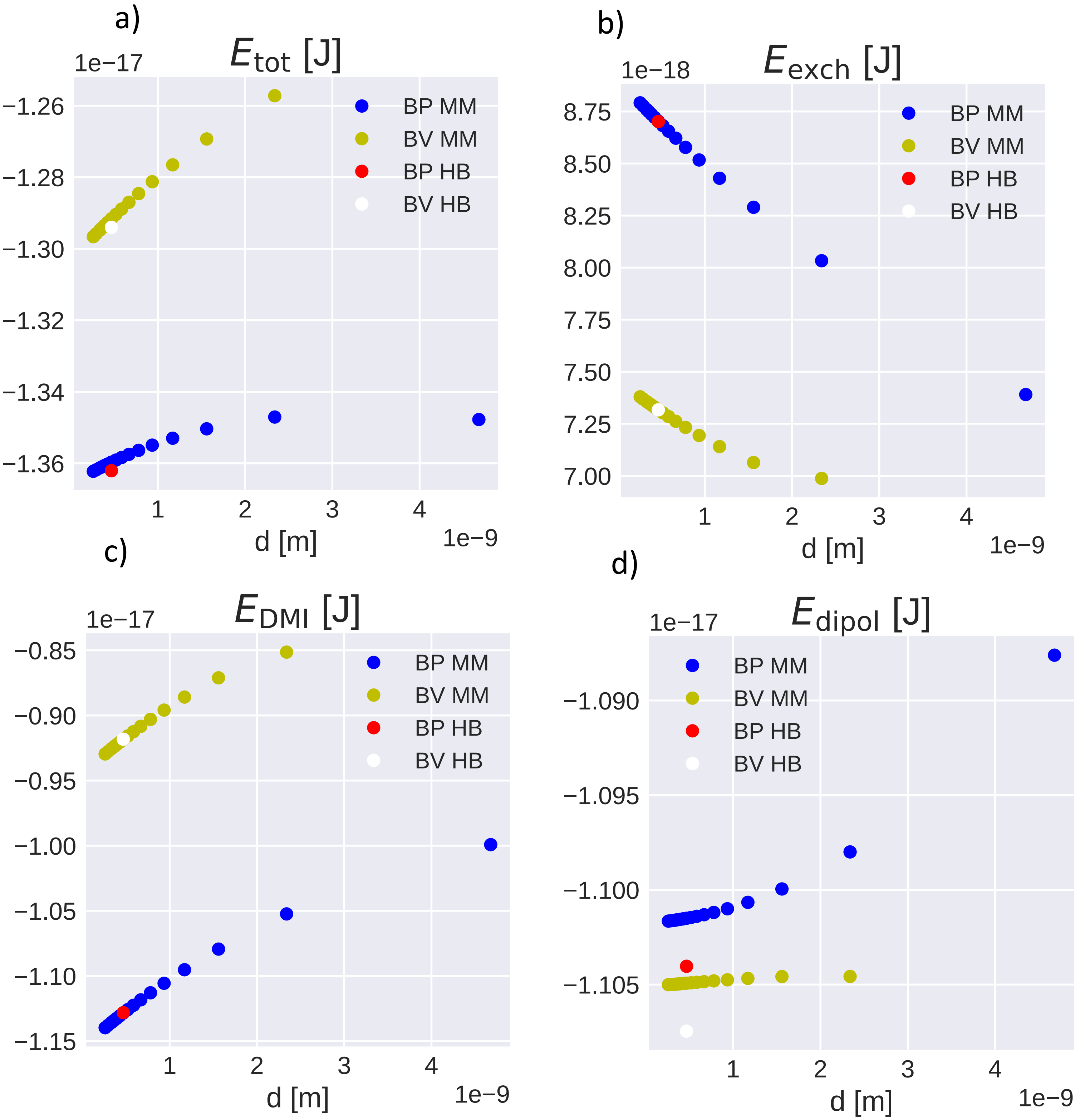

***Figure 2 a) - d):*** *total energy density and single contributions for the BP state and the BV state, for various discretizations in the MM model and the more accurate HB model. It is clearly visible that the MM energy approaches the HB value for decreasing simulation cell size.*

The trend for the single MM energies for the exchange and the DMI continues if the cell size is chosen to be smaller than the lattice constant $a$. This is not surprising, since $a$ is an arbitrarily chosen constant from a MM perspective.

To compare energies in the MM and HB models, we have subtracted the energies of the ferromagnetic state from all energy terms. The HB energies are generally in agreement with the MM model, and no qualitative differences are visible. Energy differences between micromagnetic and Heisenberg models with the same lattice spacing are tiny, suggesting that the discretization is a more important factor for the outcome than the choice of the model. For the important topic of skyrmion stability and skyrmion annihilation processes in circular disks and some related studies also the effects of the discretization have been found to be important highlighting the interest in understanding the validity of such simulations [11,27,28,40]. To corroborate the robustness of our results, we investigate the BP core magnetization in the Appendix. We especially check that the BP structure in this high DMI bilayer-systems behaves as expected for typical BP structures, even with high induced DMI [41].

## Dynamics of the stable BP

We investigate the dynamics of a stable BP configuration. We start our study from an equilibrium configuration where the BP is located at the interface between the two layers and in the middle of a disk geometry. The initial equilibrium configurations are shown in *Fig. 3.* We then create a higher energy configuration by applying an external field of $H_{\mathrm{ext,x}} = 25\ \mathrm{mT}$ which is applied in the x direction. This applied field moves the BP from the initial equilibrium position at the center of the disk to a new location in the x-y plane. When this higher energy configuration has been obtained numerically, we set the external field to zero. At this point, we have changed the energy surface, and the magnetization configuration wants to change back into the initial equilibrium configuration. We simulate the time dependence of this process and thus investigate the dynamics of the BP moving back to the center position (which is the energy minimum). We make a snapshot of the configuration every 20 ps. We further investigate how the discretization in the MM model is influencing the BP movement.

To put the observed BP dynamics into context, we repeat the process of field-shifting, as described above for the BP configuration, with the vortex configuration in the bilayer system, and with a classical vortex (CV) in a FeGe disk (i.e., without any DMI interactions for the classical vortex), where the motion is well-known and can be described analytically. It is further of interest how the bilayer vortex behaves in comparison to a classical vortex, as the former shows a substantial change in its vortex core radius $R_C$ along the thickness. From a Thiele model perspective [26], the dissipation tensor $\mathcal{D} = -\alpha\pi M_s t/\gamma \cdot (2 + ln(r_{disk}/R_C))$ depends on the radius of the vortex core $R_C$ [42]. It might be that due to the radius variation, the precessional motion becomes disturbed.

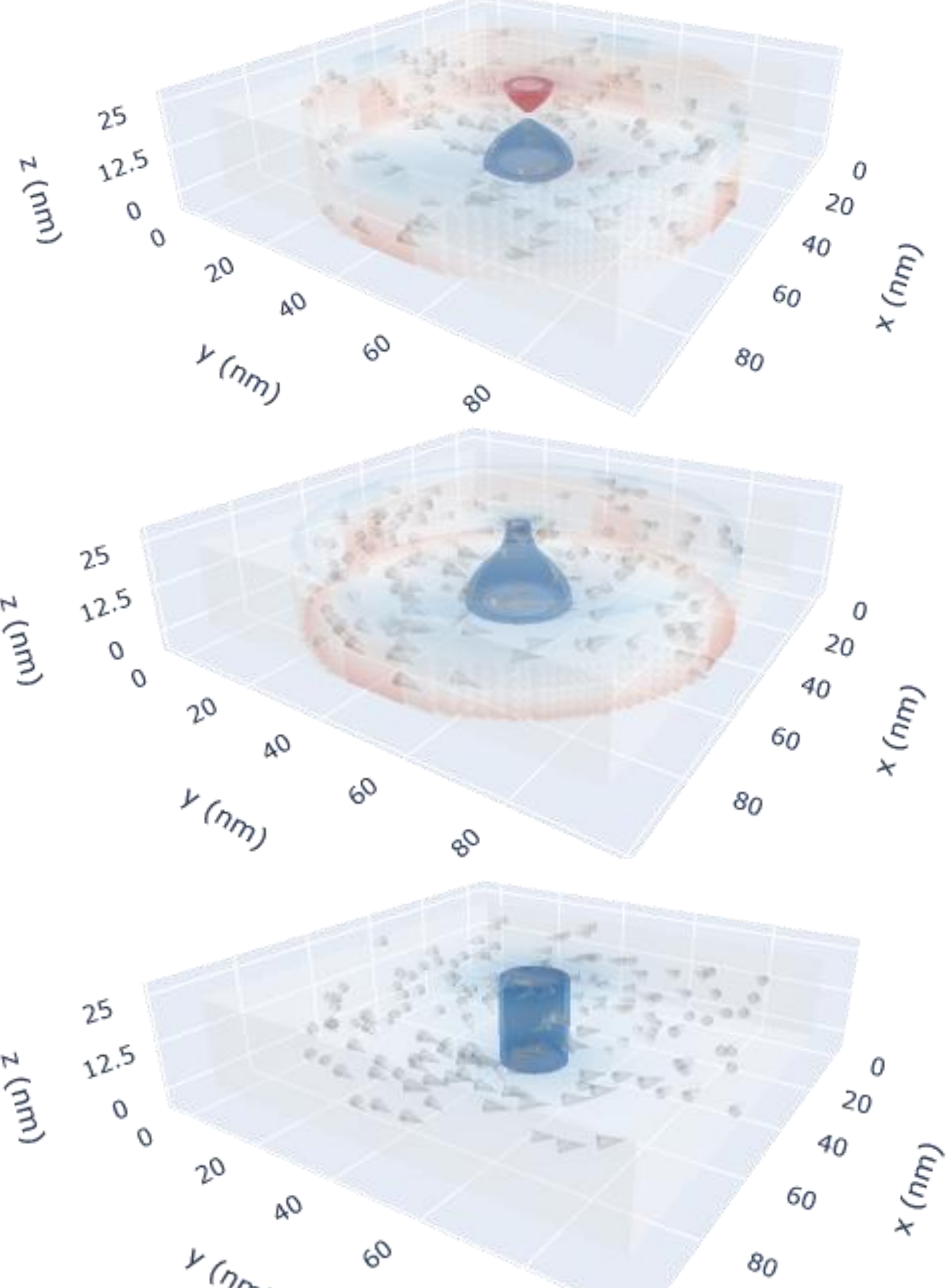

***Figure 3:*** *The three investigated structures in a cylindrical disk. The BP state in the bilayer system (top), the vortex state in the bilayer system (middle), and a classical vortex in a one-layer stack without DMI (bottom), where motion can be described analytically.*

*Fig. 3* visualizes the systems in the equilibrium state without applied field. For the classical vortex, we used a field strength of $10\ \mathrm{mT}$ to reach a similar displacement of the core compared to the systems containing DMI, which results in a large restoring force. We use circular disks for this part of the study to minimize affecting the precessional motion by a non-rotational symmetry of the sample shape. The diameter $\mathrm{d}_{\mathrm{disk}}$ of the disks is chosen with $\mathrm{d}_{\mathrm{disk}} = 200 \cdot \mathrm{a}$. The position of the vortex in one z-slice of a simulation is calculated by the center of mass of the topological charge density, calculated with the Berg-Lüscher method [43]. In more detail, an initial guess of the vortex position is given through the weighted mean of the 4 cells with the highest absolute z-magnetization, with the z-magnetization chosen as weights. A cut-off radius around this first approximation of the vortex position is chosen with $r_{\mathrm{cutoff}} = 1/4\, r_{\mathrm{disk}}$. Within this region the topological charge density is calculated. The cut-off is used to not disguise the center position by the significant edge canting that occurs in these finite-size DMI systems. For the vortex configurations, the position is calculated on an x-y data set that is obtained by averaging all vectors at that x-y position in the z-direction.
To compute the BP position, we re-use the algorithm described above for the vortex position calculation as the BP consists of one vortex in the top layer and a second vortex in the bottom layer. The BP position is defined as the mean position of these two vortex positions.

*Fig. 4* shows the movement of the BP for different discretizations of the MM mesh and within the HB model from its higher energy state back to the center of the geometry as detailed above. The center of the disk is at $y \sim 46.79\ \mathrm{nm}$. The initial points of the trajectories are found for values of $y < 37.5\ \mathrm{nm}$ due to the field acting in x direction: By moving the vortex configuration in the y-direction more of the magnetization can align with the external field. The sign of the displacement along the y-direction is determined by the circulation in the vortex configurations that make up the BP configuration. The figure shows the trajectories once the external field is switched off. We see that for the BP state, the motion shows to first approximation a straight line from the initial to the final point (note the high resolution of the x-scale). The line becomes more straight for smaller micromagnetic discretization cell sizes. For a discretization cell size identical to the one used in the Heisenberg model, the Heisenberg model trajectory and the micromagnetic trajectory coincide for a stable BP configuration. This contrasts with the case of skyrmion annihilation, where the maximum BP speed changes between the models, even at the same cell size [28].

The simulation grid of the Heisenberg model is shown in *Fig. 4* with grey grid lines. We can see that the error that is introduced by using a micromagnetic model with a lattice spacing larger than the atomistic lattice spacing $a_{\mathrm{HB}}$, that is used in the Heisenberg model, is relatively small and of the order of one lattice spacing or smaller in x-direction.

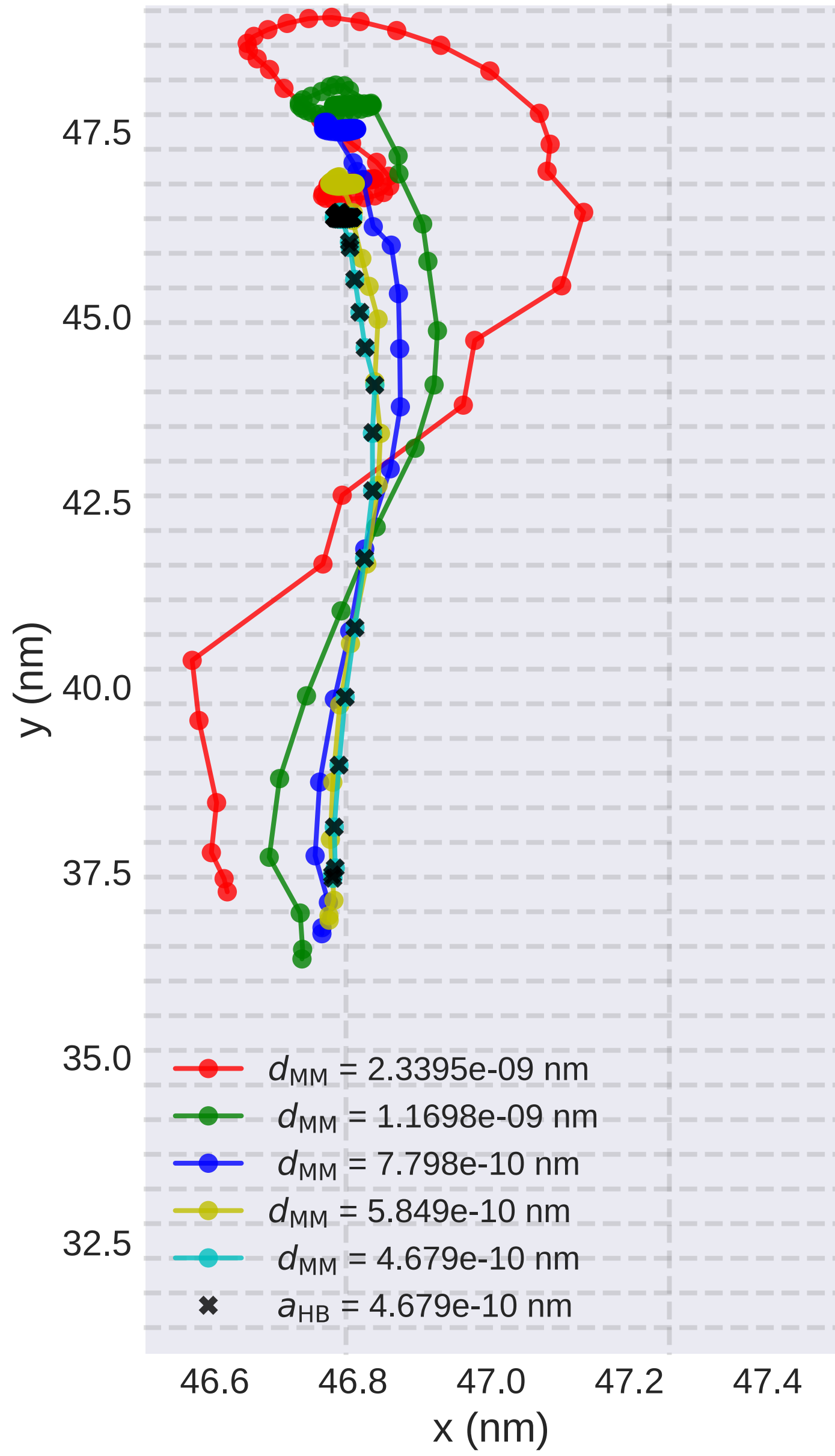

***Figure 4:*** *The motion of the BP for different cell sizes and models. The BPs are moving towards a larger y-component (starting at the bottom of the image). Every point indicates a time integration of* 20 ps. *We see that the BP is moving straight to the center of the disk after turning off the field abruptly. There is no significant precession visible. The dynamics are well described in the MM model when the cell size approaches the HB lattice constant. The underlying grid shows the HB mesh for comparison. The tracking procedure is described in the main text.*

To compare the BP trajectory with that of better-known magnetization configurations, *Fig. 5a* shows the trajectories of the BP in the bilayer system (which was already discussed in *Fig. 4*), the vortex in the bilayer system and the classical vortex (i.e., without any DMI) within the disk geometry. All results shown in this figure have been computed using the Heisenberg model, which we treat in this study as the most accurate model available. We use a maximum torque of $0.5\ \mathrm{degrees/ns}$ as the stopping criterion for the time integration.

*Fig. 5b* shows a magnified view of the different trajectories. Also, the motion of the top (green) and bottom (blue) slice of the coupled vortices in the BP state is visualized there. The classical vortex configuration (grey-blue) exhibits a precession around the center points, as expected [44]. Also, the bilayer vortex (orange) shows a smooth precession. We do not observe any perturbation by the strong vortex core variation along the thickness, probably due to the strong exchange interaction that keeps the vortex in shape.

As visible in *Fig. 5c*, the relaxation of the bilayer vortex (BV) configuration takes more time, $t_{\mathrm{relax}} \simeq 3.34\ \mathrm{ns}$, than the BP configuration ($t_{\mathrm{relax}} \simeq 1.66\mathrm{ns}$) and travels a larger distance. A large part of the distance is covered quite rapidly, while the final relaxation close to the equilibrium takes much more time due to the flat energy landscape around the equilibrium position. The classical vortex executes a similar polarization-dependent precession [45] – after the removal of an applied field of $10\ \mathrm{mT}$ – while relaxing in $4.47\ \mathrm{ns}$. The slightly different trajectories of BV and CV are a consequence of the slightly different initial positions due to different field strength required to displace the vortex in a system with/without DMI (no effort was put into finding the correct field strength for the exact same initial position as it is not required for this work). The different relaxation times might be explained by consulting the Thiele model. As can be seen in *Fig. 3*, the radius of the BP and BV spin structure changes throughout the sample thickness, which will lead to different effective core radii. Especially around the BP itself, the core radius is reduced [15]. Also the BV state relaxes faster than the classical vortex. We speculate that the additional DMI results in higher energy gain with the spatial displacement, leading to stronger relaxing forces. DMI-induced changes of vortex dynamics have already been reported [46].

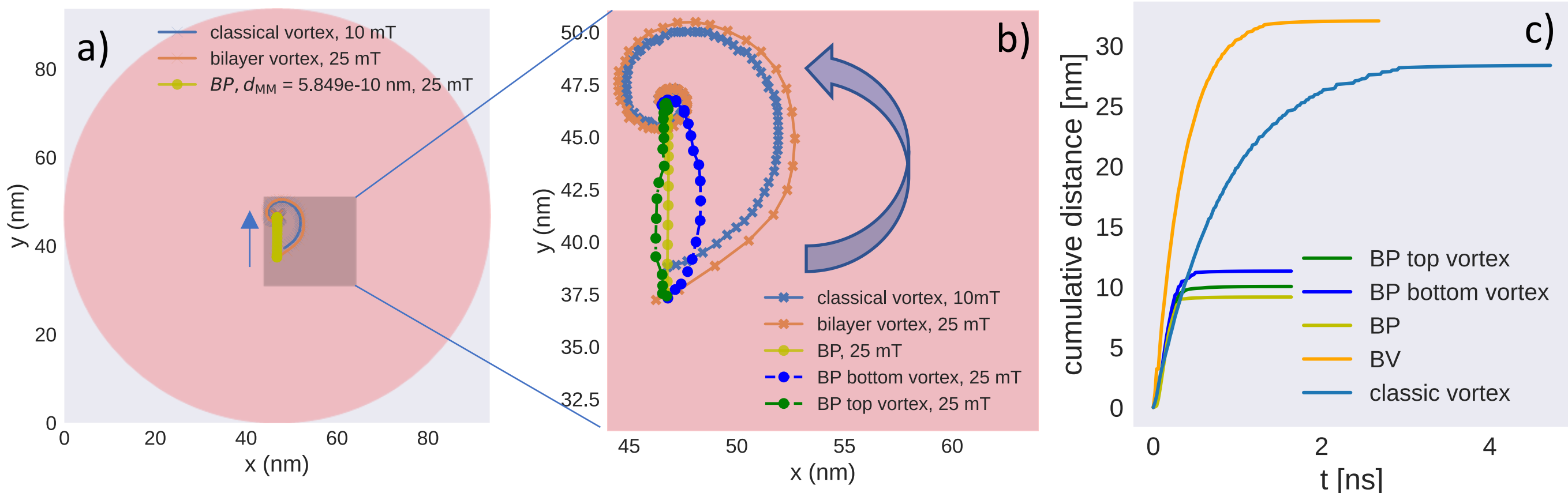

***Figure 5:** Comparison of the motion of the BP, the bilayer vortex, and the classical vortex in the HB model. Every point indicates a time step of $20\ ps$. The vortices show typical precession around the center point, while the BP trajectory leads straight to the center. Blue arrows indicate the direction of motion.* ***a)*** *The whole cylindrical disk is shown for a good view of the scales of the movement.* ***b)*** *Zoom into the central region to better see the trajectories. Here, we have added the trajectory of the vortex configuration in the bottom layer (blue) that represents the bottom part of the BP configuration and separately the trajectory of the vortex configuration in the top layer (green) that represents the top part of the BP configuration. We see that they follow a highly damped precession, while the precessional damping is smaller for the larger of the coupled vortices.* ***c)*** *The cumulative travelled distance of the investigated quasi-particles. The vortices in the bilayer system relax much faster into the equilibrium state than the classical vortex and experiences higher velocities.*

The motion of the two vortices (one in each layer) that form the BP is shown in green (BP top vortex) and blue (BP bottom vortex) in *Fig. 5b* and *c*. Both vortices show a precession in opposite directions due to their opposite polarization. However, the precession is significantly damped (in comparison to the vortex configurations) due to the coupling of the two vortices that want to precess in opposite directions at the layer interface. The bottom layer shows a larger amount of precession motion;

presumably, this can be attributed to the bottom layer being thicker than the top layer, and thus giving the vortex configuration more freedom to deviate from the configuration at the layer interface. Next, we check if the precession of two vortices in opposite directions can rupture the Bloch point configuration. We have calculated the largest distance between the vortex core position in the lowest discretization plane in the top layer (i.e., just above the layer interface) and the vortex core position in the highest discretization plane (i.e., just below the layer interface) and find this to be $0.173\ \mathrm{nm}$ in the HB model. So, the vortices can be considered to stay coupled across the layer interface during the relaxation. The deviation observed in *Fig. 5b* originates from a deformation of the vortices further away from the layer interface. The BP trajectory (olive green in *Fig. 5b*) appears as a straight line although the two coupled vortices have different lengths in our system (ratio 3/2, given by the relative thickness of the two stacks). The BP motion is thus for our system not influenced significantly by the two vortex tubes of different lengths. Simulation scripts to repeat the study are provided as open access [47,48].

## Conclusion

We have shown, using a Heisenberg model, that the BP configuration has a lower energy than a vortex state in the bilayer system, and is a stable configuration. The vortex state has higher energy since the DMI energy can only be minimized for one of the two layers. The micromagnetic framework can yield (at least qualitatively) correct results for the Bloch point system studied here. With finer discretization of the finite-difference mesh, the micromagnetic results approach the gold standard results of the HB model quantitatively.

The energies predicted by the micromagnetic model depend weakly on the lattice discretization used. This is similarly the case for the Bloch point system in a DMI material but also for a vortex configuration in a ferromagnetic material – one of the areas in which micromagnetic simulations are traditionally used as the workhorse of computational support for research and development. The energies predicted by the micromagnetic model when using the same lattice spacing as in the Heisenberg model are very close to the energies obtained from the Heisenberg model.

While studying the dynamics of the BP, we find that the motion of the BP is along a straight line whereas a vortex configuration shows precessional motion. The BP is composed of two vortices with identical circulation and opposite polarization. The BP configuration and the exchange coupling of the two vortices across the bilayer interface results in the straight trajectory where the opposite expected motion in the individual layers cancels out. The Bloch point also does not overshoot as known for vortex systems due to their precessional motion. Fast relaxation of solitons may be an important feature for potential applications in storage and computing devices.

## Acknowledgment

The group in Mainz acknowledges funding by the emergent AI center, funded by Carl-Zeiss Stiftung, and by and the German Research Foundation (DFG SFB TRR 173 project A01 and B02, 49741853 and 268565370, SPIN+X). This work was financially supported by the EPSRC Programme grant on Skyrmionics (EP/N032128/1).

## Authors contributions

Thomas Brian Winkler performed the study and wrote the manuscript. M.B. and H.F. conceived the study. M.K. and H.F. supervised the study. All authors contributed to the project with discussions and contributed to the manuscript.